\documentclass[twocolumn,floats,prb,aps,showkeys]{revtex4}

\usepackage{graphicx}
\usepackage{dcolumn} 
\usepackage{amsmath}
\usepackage{exscale}
\usepackage{bm}      
\usepackage{color}
\usepackage{epsfig,psfrag,subfigure,subeqnarray}
\usepackage{bbm}
\usepackage{amsbsy}
\usepackage{amssymb}
\usepackage{enumitem}
\usepackage{mathrsfs}

\def\ran{\rangle}

\def\vk{{\bf k}}

\def\vr{{\bf r}}
\def\vu{{\bf u}}

\def\vp{{\bf p}}

\def\v0{{\bf 0}}
\newcommand{\bd}{\begin{equation}}
\newcommand{\ed}{\end{equation}}
\newcommand{\be}{\begin{equation}}
\newcommand{\ee}{\end{equation}}
\newcommand{\bt}{\begin{split}}
\newcommand{\et}{\end{split}}

\newcommand{\bn}{\begin{align}}
\newcommand{\en}{\end{align}}

\newcommand{\bea}{\begin{eqnarray}}
\newcommand{\eea}{\end{eqnarray}}
\newcommand{\ba}{\begin{array}}
\newcommand{\ea}{\end{array}}
\newcommand{\nn}{\nonumber}

\begin{document}

\title{Missing understanding of the phase factor between  valence-electron and hole operators}

\author{Shiue-Yuan Shiau$^1$, and Monique Combescot$^2$}\
\affiliation{(1) Physics Division, National Center for Theoretical Sciences, Hsinchu, 30013, Taiwan}
\affiliation{(2) Sorbonne Universit\'e, CNRS, Institut des NanoSciences de Paris, 75005 Paris, France}

\begin{abstract}
This paper provides the long-missing foundation to connect semiconductor and  atomic notations and to support results incorrectly obtained  by doing as if semiconductor electrons possessed an orbital angular momentum. We here show that the phase factor between valence-electron destruction operator and hole creation operator is the same as the one between particle and antiparticle in quantum relativity, namely $\hat{a}_{m}=(-1)^{j-m} \hat{b}^\dag_{-m}$ provided that $m=(j,j-1\cdots,-j)$ labels  the degenerate states of the $(2j+1)$-fold electron level at hand. This result is remarkable because $(i)$ the hole is definitely not a naive antiparticle due to the remaining valence  electrons; $(ii)$ unlike atomic electrons in a central potential, semiconductor electrons in a periodic crystal do not have \textit{orbital angular} momentum $\textbf{L}=\vr\wedge\vp$ nor \textit{angular} momentum $\textbf{J}=\textbf{L}+\textbf{S}$. Consequently, $(j,m)$ for semiconductor electrons merely are convenient notations to label the states of a degenerate level. To illustrate the physical implications, we discuss the interband couplings between photons and semiconductor, in terms of valence electrons and  of holes: the phase factor is crucial to establish that bright excitons are in a spin-singlet state.
\end{abstract}

\date{\today}
\maketitle

\section{Introduction\label{sec1}}

Valence hole is a major concept of semiconductor physics\cite{Kittelbook,Merminbook,Cardona}. Thinking in terms of holes allows us to forget all the electrons that remain in the valence band, with just one requirement: we must keep in mind that the valence band is still there and can ``boil'' through the virtual excitations of electron-hole pairs, as in the microscopic processes responsible for the semiconductor dielectric constant\cite{Monicbook}.\

In a phenomenological approach to semiconductors, one says that removing an electron with momentum $\vk$ and spin $s_z$ from the valence band, leaves this band full with a hole having a momentum $-\vk$ and spin $-s_z$; so, the destruction of a $(\vk,s_z)$ valence electron corresponds to the creation of a $(-\vk,-s_z)$ hole. Although it is commonly mentioned that the upper valence level is spatially degenerate, little is said\cite{Hartmutbook} about the relation between the spatial indices of  valence-electrons and holes, and very rarely is the existence of a phase factor between their destruction and creation operators  mentioned. A possible reason is that, in most physical effects, the number of electrons in each band stays constant because changing this number requires a large energy, of the order of the band gap. Yet, this number does change in a set of effects of  technological interest like the ones associated with semiconductor-photon interaction, that is, electron excitation or de-excitation through photon absorption or emission. Then, this phase factor is required to properly derive polarization effects\cite{Monicprb1990}.

Here we show that a phase factor does exist between destruction and creation operators of a semiconductor electron in a periodic crystal. By labeling the states of a ($2j+1$)-fold level as $m = (j,j-1\cdots,-j)$, this phase factor reads as  
\be
\label{1}
\hat{a}_{m}=(-1)^{j-m} \hat{b}^\dag_{-m}\,.
\ee

\begin{figure}[t]
\begin{center}
\includegraphics[trim=3.5cm 4cm 4cm 4cm,clip,width=2.8in] {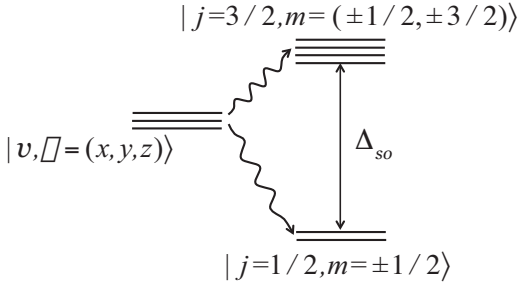}
\end{center}
\vspace{-0.5cm}
\caption{\small Spin-orbit interaction splits the 6-fold valence states $|v,\mu\ran\otimes |\pm1/2\ran$ into the 4-fold upper subband with states labeled as $m=(\pm1/2,\pm3/2)$ and the 2-fold lower subband with states labeled as  $m=\pm1/2$.}
\label{fig:1}
\end{figure}

We will prove  this result in a few physically relevant cases: (\textit{i}) the  2-fold level of a spin state, but also of a 2-fold level resulting from spin-orbit interaction in a 3-fold spatial level, as shown in Fig.~\ref{fig:1}, (\textit{ii}) the 3-fold spatial level of the upper valence band, for states labeled as $(\pm 1,0)$ instead of (\textit{x,y,z}), and (\textit{iii}) the 4-fold level of the upper spin-orbit subband, for states labeled as $(\pm 3/2,\pm 1/2)$.

Before going further, we wish to stress that semiconductor physicists have used atomic notations for quite a long time to describe  semiconductor energy levels, without questioning the validity of the physics behind these notations. These atomic notations actually lead to incorrect understandings when taken seriously. Indeed, in an hydrogen atom, the Coulomb potential experienced by the electron has a spherical symmetry\cite{cohenbook}. Due to the  orbital angular momentum $\textbf{L}=\vr\wedge\vp$, the electron orbital states have a $(2l+1)$ degeneracy with a $(-1)^l$ parity. Moreover, the spin-orbit interaction for atomic electrons can be readily written as $\textbf{L}\cdot\textbf{S}$, from which the spin-orbit eigenstates are easy to derive by introducing the angular momentum $\textbf{J}=\textbf{L}+\textbf{S}$. 

By contrast, electrons in a crystal feel a periodic potential. As a result, they do not possess an \textit{orbital angular} momentum $\textbf{L}$. The strongest reason is that while $\vr$ is a finite quantity for bound atomic electrons, it is infinite for itinerant electrons in a crystal. Another reason is the physical misconception it invokes: indeed, the lowest conduction state and the highest valence states are commonly called S and P; they  respectively are 1-fold and 3-fold like $l=(0,1)$ atomic states, but their parity is not $(-1)^l$: in crystals with inversion symmetry, all valence states are even and all conduction states are odd. Since electrons in a periodic crystal do not have orbital angular momentum, they cannot have an angular momentum $\textbf{J}$. As a direct consequence, the spin-orbit interaction for semiconductors does not read as $\textbf{L}\cdot\textbf{S}$. This  makes their spin-orbit eigenstates more tricky to derive. Until very recently, the only correct derivation of these eigenstates uses the double-group symmetries of the group theory\cite{Falicov,Dresselhaus,Ivchenko,Fishman}. We have recently shown\cite{MonicPRB2019} how to obtain the spin-orbit eigenstates for semiconductor crystals without using this nice but heavy and poorly known formalism. Our main conclusion is that  semiconductor spin-orbit eigenstates indeed have the same form as their atomic counterparts, regardless of the  parity of the spatial states, even or odd. This supports using atomic notations  to label degenerate states in cubic crystals. Yet, the fact that they do not correspond to the same physics is a real danger. More generally, one should be very careful with semiconductor results obtained from the atomic context\cite{Cardona}, especially when dealing with quantities as subtle as  phase factors.

\section{2-fold spin level}

A 2-fold degeneracy corresponds to $2j+1=2$, that is, $j=1/2$, its two states being labeled as $m=\pm1/2$. Let us here rederive Eq.~(\ref{1}) in the case of spin. Equation (\ref{1}) gives the link between the destruction operator of an electron with spin ($\pm 1/2$) and the creation operator of a hole with spin ($\mp 1/2$), as 
\be
\label{5}
\hat{a}_{\pm \frac{1}{2}}=\pm \,\hat{b}^\dag_{\mp\frac{1}{2}}\,.
\ee

To prove it  from scratch, we can consider a spin along an $\vu$ direction with Euler angles $(\theta,\varphi)$ in the $(x,y,z)$ frame. It is represented by the  $2\times2$ matrix  
\bea
\hat{S}_\vu &=&\sin\theta ( \cos \varphi \hat{S}_x+  \sin \varphi \hat{S}_y )+\cos\theta \hat{S}_z
\nn
\\
&=& \frac{\hbar}{2} \left( \begin{array}{cc}
\cos \theta & e^{-i\varphi}\sin\theta \\ 
e^{i\varphi}\sin\theta & -\cos \theta
\end{array}
\right)\,,\label{Svu6}
\eea
as obtained by using the Pauli matrices written in the $|\pm 1/2\ran_z$ basis of the $\hat{S}_z$ eigenstates, namely,
\be
\hat{S}_z\left|\pm \frac 1 2\right\ran_z =\pm \frac{\hbar}{2} \left|\pm \frac 1 2\right\ran_z\,.
\ee

From Eq.~(\ref{Svu6}), we can deduce the $\hat{S}_\vu$ eigenstates
\be
\hat{S}_\vu \left|\pm \frac 1 2\right\ran_\vu =\pm \frac{\hbar}{2} \left|\pm \frac 1 2\right\ran_\vu
\ee
 in the $|\pm 1/2\ran_z$ basis as
\be
\left|\pm \frac 1 2\right\ran_\vu =\cos \frac \theta 2 \left|\pm \frac 1 2\right\ran_z  \pm  e^{\pm i  \varphi }\sin\frac \theta 2 \left|\mp \frac 1 2\right\ran_z\,,\label{app:etoh:pm12u}
\ee
with an overall phase factor that must be taken equal to 1 to recover the $\theta\rightarrow 0$ limit.

Let us now introduce the creation operators for the $\hat{S}_z$ and $\hat{S}_\vu$ eigenstates, defined as
\be
\left|\pm \frac 1 2\right\ran_z= \hat{a}^\dag_{(\pm\frac 1 2)_z}|v\ran
\qquad \left|\pm \frac 1 2\right\ran_\vu= \hat{a}^\dag_{(\pm\frac 1 2)_\vu}|v\ran\,,
\ee
where $|v\ran$ denotes the vacuum state. Equation (\ref{app:etoh:pm12u}) gives their link as 
\be
\hat{a}^\dag_{(\pm\frac 1 2)_\vu} = \cos \frac \theta 2~ \hat{a}^\dag_{(\pm\frac 1 2)_z} \pm e^{\pm i  \varphi }\sin\frac \theta 2 ~\hat{a}^\dag_{(\mp\frac 1 2)_z}\,.\label{app:etoh:a+pm12u}
\ee

 To go further, we note that the phase factor between electron and hole operators must be the same, whatever the $(x,y,z)$ axes, that is, whatever $\vu$. This implies
\begin{subeqnarray}
\hat{a}_{(\pm\frac 1 2)_z}&=& e^{ig_\pm}\hat{b}^\dag_{(\mp\frac 1 2)_z}\,,\\
\hat{a}_{(\pm\frac 1 2)_\vu}&=& e^{ig_\pm}\hat{b}^\dag_{(\mp\frac 1 2)_\vu}\,.
\end{subeqnarray}
Equation (\ref{app:etoh:a+pm12u}) then gives 
\bea
\hat{a}_{(\pm\frac 1 2)_\vu}&=&\Big ( \cos \frac \theta 2~ \hat{a}^\dag_{(\pm\frac 1 2)_z} \pm e^{\pm i  \varphi }\sin\frac \theta 2 ~\hat{a}^\dag_{(\mp\frac 1 2)_z}\Big)^\dag  \\
&=& \cos \frac \theta 2 \Big (e^{ig_\pm}~ \hat{b}^\dag_{(\mp\frac 1 2)_z}\Big) \pm e^{\mp i  \varphi }\sin\frac \theta 2 \Big (
e^{ig_\mp}~ \hat{b}^\dag_{(\pm\frac 1 2)_z}\Big)\,,
\nn
\eea 
which  leads to $\hat{a}_{(\pm\frac 1 2)_\vu}=e^{ig_\pm}\hat{b}^\dag_{(\mp\frac 1 2)_\vu}$ whatever $( \theta,\varphi)$, provided that $e^{ig_\pm}=-e^{ig_\mp}$, in agreement with the general form (\ref{1}), namely
\be
\label{15}
\hat{a}_m= (-1)^{\frac 1 2-m}     \hat{b}^\dag_{-m} \,\,\,\,\,\,\,\,\,\,\,\textrm{for} \,\,\,\,\,\,\,\,\, m=\pm \frac 1 2 \,.
\ee
This phase factor ensures that, when turning from electron destruction operator to hole creation operator, the hole spin transforms under a rotation in the same way as the electron spin.

\section{3-fold spatial level}

Electrons in a semiconductor are controlled by the crystal symmetry.

\subsection{States labeled as $\mu=(x,y,z)$}
A way to label the electron states of a 3-fold level  in a cubic crystal is using its crystal axes $(x,y,z)$. Due to cyclic symmetry, they  play the same role; so, we must have
\be
\hat{a}_\mu=\hat{b}^\dag_\mu  \,\,\,\,\, \,\,\,\,\,\,\textrm{for} \,\,\,\,\,\,\,\,\,      \mu=(x,y,z)\,.
\ee

\subsection{ States labeled as $m=(\pm1,0)$}

\noindent $\bullet$ By using the Landau-Lifshitz phase factor\cite{Landau} for spherical harmonics
\begin{subeqnarray}
Y_{1,\pm1}&=& \mp i \sqrt{\frac{3}{8\pi}} \sin \theta e^{\pm i\varphi} =\sqrt{\frac{3}{4\pi}} \,\,\frac{\mp ix +y}{r\sqrt{2}}\,,\\
Y_{1,0}&=& i \sqrt{\frac{3}{4\pi}} \cos \theta = \sqrt{\frac{3}{4\pi}} \,\, \frac{iz}{r}\,,
\end{subeqnarray}
that fulfill $Y_{\ell,-\ell}=Y^*_{\ell,\ell}$, as required by the particle-antiparticle quantum symmetry\cite{Berestetskii}, we are led to write the creation operators for  3-fold  spatial states labeled as $m=(\pm1,0)$ in terms of the creation operators for states labeled as $\mu=(x,y,z)$, as
\be
\label{2}
\hat{a}^\dag_{\pm 1}=\frac{\mp i\, \hat{a}^\dag_x +\hat{a}^\dag_y}{\sqrt{2}} \qquad  \hat{a}^\dag_{0}= i \hat{a}^\dag_z\,.
\ee
Since $\hat{a}_\mu=\hat{b}^\dag_\mu$, this readily gives
\bea
\label{15}
\hat{a}_{\pm1}&=& \frac{\pm i \hat{a}_x +\hat{a}_y}{\sqrt{2}}=\frac{\pm i \hat{b}^\dag_x+\hat{b}^\dag_y}{\sqrt{2}}=\hat{b}^\dag_{\mp1}\,,\label{ap_eth_21a}\\
\hat{a}_{0}&=& -i \hat{a}_z=-i\hat{b}^\dag_z=-\hat{b}^\dag_{0}\,,\label{ap_eth_21b}
\eea
which agrees with the general link 
\be
\label{27}
\hat{a}_{m}=(-1)^{1-m} \hat{b}^\dag_{-m} \,\,\,\,\,\,   \,\,\,\,\textrm{for} \,\,\,\,\,\,\,\,\,            m=(\pm1,0)\,,
\ee 
since 3-fold degeneracy corresponds to $2j+1=3$, that is, $j=1$.

\noindent $\bullet$ We wish to mention that by adopting the commonly used phase factor for spherical harmonics (see, e.g., Refs. [\onlinecite{Cardona,cohenbook}] without the $i$ phase), namely
\begin{subeqnarray}
\widetilde{Y}_{1,\pm1}&=& \mp  \sqrt{\frac{3}{8\pi}} \,\sin \theta e^{\pm i\varphi} =\sqrt{\frac{3}{4\pi}}\, \frac{\mp x -iy}{r\sqrt{2}}\,,\\
\widetilde{Y}_{1,0}&=&  \sqrt{\frac{3}{4\pi}}\, \cos \theta = \sqrt{\frac{3}{4\pi}}\,\,   \frac{z}{r}\,,
\end{subeqnarray}
we would be led to take
\bea
\hat{\tilde{a}}^\dag_{\pm1}= \frac{\mp \hat{\tilde{a}}^\dag_x -i\hat{\tilde{a}}^\dag_y}{\sqrt{2}}\,
\,\,\,\,\,\,\,\,\,\,\,\,\,\,\,\,\,\,\,\,\,\,\,\,
\hat{\tilde{a}}^\dag_{0}= \hat{\tilde{a}}^\dag_z\,,
\eea
instead of Eq.~(\ref{2}), from which we would find
\bea
\hat{\tilde{a}}_{\pm1}&=& \frac{\mp\hat{\tilde{a}}_x +i\hat{\tilde{a}}_y}{\sqrt{2}}=\frac{\mp \hat{\tilde{b}}^\dag_x+i\hat{\tilde{b}}^\dag_y}{\sqrt{2}}=-\hat{\tilde{b}}^\dag_{\mp1}\,,\\
\hat{\tilde{a}}_{0}&=&\hat{\tilde{a}}_z=\hat{\tilde{b}}^\dag_z=\hat{\tilde{b}}^\dag_{0}\,,
\eea
that is, $\hat{\tilde{a}}_{m} =-(-1)^{1-m} \hat{\tilde{b}}^\dag_{-m}$, which differs from Eq.~(\ref{1}) by a  minus sign. This would force us to take a different  phase factor in front of the RHS of Eq.~(\ref{1}) for $j=1$, without any physical reason.

\noindent $\bullet$ This shows that in order for $\hat{a}_{m}$ to fulfill the general link (\ref{1}), it is necessary to adopt a phase factor that preserves the fundamental symmetry between particle and antiparticle.

 Let us note that, by consistency, this imposes to take the  creation operators and polarization vectors of circularly polarized photons propagating along $\textbf{e}_z$ as
\bea
\label{22}
\hat{\alpha}^\dag_{\pm}=\frac{\mp i \,\alpha^\dag_x +\alpha^\dag_y}{\sqrt{2}} \,,
\\
\textbf{e}_{\pm }=\frac{\mp i \,\textbf{e}_x +\textbf{e}_y}{\sqrt{2}} \,,
\eea
in order to have 
\bea
\label{23'}
\sum_{\mu=(x,y)}  \textbf{e}_\mu \, \alpha_\mu =\sum_{\ell=\pm}
\textbf{e}_\ell\,
\alpha_\ell \,,
\eea
as necessary to write the field vector potential in terms of linearly or circularly polarized photons.

\section{4-fold spin-orbit level}

We now consider a 4-fold level with states labeled as $m=(\pm3/2,\pm1/2)$,  as the one resulting from spin-orbit interaction in a 3-fold spatial level (see Fig.~\ref{fig:1}).

It has been shown\cite{MonicPRB2019} that the creation operators for electrons in this 4-fold level read in terms of creation operators $\hat{a}^\dag_{m;s_z}$ for electrons with spin $s_z=\pm 1/2$ in a spatial state $m=(\pm1,0)$, as 
\bea
\hat{a}^\dag_{\pm \frac{3}{2}} &=&\hat{a}^\dag_{\pm 1; \pm\frac 1 2}\,,\label{app:etoh:a+323eta2=}
\\
\hat{a}^\dag_{\pm\frac{1}{2}}&=&\frac{\hat{a}^\dag_{\pm1; \mp \frac{1}{2}}+ \sqrt{2}\,\hat{a}^\dag_{0; \pm \frac 1 2}  }{\sqrt{3}}\,.\label{app:etoh:a+123eta2=}
\eea

Equations (\ref{15}) and (\ref{27}) provide the link between the destruction operator of a $(m,s_z)$ electron and the creation operator of its $(-m,-s_z)$ hole as
\be
\label{36}
\hat{a}_{m;\pm\frac{1}{2}}=\pm (-1)^{1-m} \, \hat{b}^\dag_{-m;\mp\frac{1}{2}}\,.
\ee
By using the above result into Eqs.~(\ref{app:etoh:a+323eta2=},\ref{app:etoh:a+123eta2=}), we obtain the expected sign change between $\hat{a}_{m}$ and $\hat{b}^\dag_{-m}$ when $m$ varies, namely  
\bea
\label{37}
\hat{a}_{\pm \frac{3}{2}} &=&\hat{a}_{\pm1,\pm\frac 1 2}=\pm \hat{b}^\dag_{\mp 1; \mp\frac 1 2}=\pm\hat{b}^\dag_{\mp \frac{3}{2}}\,,\\
\hat{a}_{ \pm\frac{1}{2}}&=&\mp \frac{\hat{b}^\dag_{\mp1; \pm \frac{1}{2}}+ \sqrt{2}\,\hat{b}^\dag_{0; \mp \frac 1 2}  }{\sqrt{3}}=\mp  \hat{b}^\dag_{\mp\frac{1}{2}}\,.
\eea
This proves that Eq.~(\ref{1}) is also valid for 4-fold spin-orbit level, 
\be
\label{33}
\hat{a}_{m}=(-1)^{\frac{3}{2}-m} \hat{b}^\dag_{-m} \,\,\,\,\,\,   \,\,\,\,\textrm{for} \,\,\,\,\,\,\,\,\,            m=\left(\pm\frac{3}{2},\pm\frac{1}{2}\right)\,,
\ee
since a 4-fold degeneracy corresponds to $2j+1=4$, that is, $j=3/2$.

\section{2-fold spin-orbit level}

As a last physically relevant example in semiconductor physics, we consider the 2-fold spin-orbit level. This degeneracy corresponds to $j=1/2$ with states labeled as $m=\pm 1/2$ like spin states (see Fig.~\ref{fig:1}). The creation operators of this 2-fold level, that are orthogonal to the $\pm 1/2$ states of the 4-fold spin-orbit level, read as
\be
\hat{a}'^\dag_{ \pm \frac{1}{2}}=\pm\frac{\sqrt{2}~ \hat{a}^\dag_{\pm 1; \mp\frac{1}{2}}-\hat{a}^\dag_{0;\pm\frac 1 2}}{\sqrt{3}}\,.
\ee
 Using Eq.~(\ref{36}), we then get 
\bea
\hat{a}'_{ \pm \frac{1}{2}}&=&\pm \frac{\sqrt{2}~ \hat{a}_{\pm1; \mp\frac{1}{2}}-\hat{a}_{0;\pm\frac 1 2}}{\sqrt{3}}\nn\\
&=&\pm \frac{(\mp)\sqrt{2}~ \hat{b}^\dag_{\mp1; \pm\frac{1}{2}}+(\pm) \hat{b}^\dag_{0;\mp\frac 1 2}}{\sqrt{3}}\,,
\eea
which reduces to 
\be
\hat{a}'_{ \pm \frac{1}{2}}=-\frac{\sqrt{2}~ \hat{b}^\dag_{\mp1; \pm\frac{1}{2}}- \hat{b}^\dag_{0;\mp\frac 1 2}}{\sqrt{3}}=\pm \hat{b}'^\dag_{\mp\frac{1}{2}}\,.
\ee
This equation again fulfills the same equation (\ref{15}) valid for 2-fold degeneracy.

\section{Photon-semiconductor coupling}

To illustrate how the $(j,m)$ indices and the phase factor that appear when turning electron into hole, have a real impact on physics,  we consider the part of photon-electron coupling that is associated with interband processes. For cubic semiconductors, this coupling has the following form\cite{Monicbook} due to cubic symmetry
\be\label{pe1}
\sum_{\mu=(x,y)}\alpha^\dag_\mu \left(P^* a^\dag_{v,\mu}a_c + P a^\dag_c a_{v,\mu} \right) +h.c.\,,
\ee
where $\alpha_\mu^\dag$ creates a photon that propagates along $\textbf{e}_z$ and has a polarization vector $\textbf{e}_\mu$ along the cubic axes  $\mu=(x,y)$, 
$a^\dag_c$ creates a nondegenerate conduction electron and $a^\dag_{v,\mu}$ creates a
degenerate valence electron with spatial index $\mu$.

$\bullet$ Using Eqs.~(\ref{15},\ref{22}), it is easy to check that
\bea
\label{pe2}
\sum_{\mu=(x,y)}\alpha^\dag_\mu a^\dag_{v,\mu}&=&\sum_{\ell=\pm}\alpha^\dag_\ell a^\dag_{v, -\ell}\,,
\\
 \sum_{\mu=(x,y)}\alpha^\dag_\mu a_{v,\mu}&=&\sum_{\ell=\pm}\alpha^\dag_\ell a_{v,\ell}\,.
\eea 
So, we can rewrite the interband photon-electron coupling as
\be\label{pe3}
\sum_{\ell=\pm}\alpha^\dag_\ell \left(P^* a^\dag_{v,-\ell}a_c + P a^\dag_c a_{v,\ell} \right) +h.c.\,.
\ee
This shows that the creation of a photon with polarization vector $\textbf{e}_\ell$ goes along with the creation of a valence electron with spatial index $-\ell$ and the destruction of a valence electron with spatial index $\ell$: the total $\ell$ is conserved in the coupling.

$\bullet$ By turning from conduction and valence electrons to electron and hole in Eq.~(\ref{pe3}), we find,
using Eq.~(\ref{27}), 
\be\label{pe4}
\sum_{\ell=\pm}\alpha^\dag_\ell \left(P^* b_{\ell}\,a + P a^\dag b^\dag_{-\ell} \right) +h.c.\,.
\ee
This again shows that the total $\ell$ is conserved in the coupling: a photon with circular polarization $\ell$ goes along with the destruction of a pair having a $\ell$ hole or the creation of a pair having a $-\ell$ hole.

$\bullet$ Let us now introduce the spin. Since photons  do not act on spin, the electron in the photon-electron coupling does not change its spin. So, Eq.~(\ref{pe3}) becomes
\be\label{pe5}
\sum_{\ell=\pm}\alpha^\dag_\ell \sum_{s=\pm1/2} \left(P^* a^\dag_{v,-\ell,s}a_{c,s} + P a^\dag_{c,s} a_{v,\ell,s} \right) +h.c.\,.
\ee
By turning valence electron into hole, according to Eq.~(\ref{36}), we get
\be
\label{pe6}
\sum_{\ell=\pm}\alpha^\dag_\ell \!\sum_{s=\pm1/2}\!\!(-1)^{\frac{1}{2}-s} \left(P^* b_{\ell,-s}a_{s} {+} P a^\dag_{s} b^\dag_{-\ell,-s} \right) +h.c.\,.
\ee
In addition to again having the same correlation between spatial indices $\ell$, the above equation interestingly shows that the phase factor rule brings a spin-dependent minus sign which determines that the pair coupled to photon is not only made from opposite-spin carriers, but this pair must be in the spin-singlet combination, namely
\be
 B^\dag_{-\ell,S=0}=\frac{a^\dag_{1/2} b^\dag_{-\ell,-1/2}- a^\dag_{-1/2} b^\dag_{-\ell,1/2}}{\sqrt{2}}\,.
\ee 
The excitons  with electron-hole pairs in a spin-singlet state are called ``bright'' because they are the ones coupled to photons.  By contrast, the excitons made of spin-triplet electron-hole pairs, even when their carriers have opposite spins, like the triplet $(S=1, S_z=0)$,  are called ``dark'' because they are not coupled to photons.

\section{Conclusion}

This work presents the missing  foundation to extend atomic notations to semiconductors, provided that they are understood differently.

 In the semiconductor context, these notations  only constitute  a convenient way to label the  states of a degenerate level: indeed, the degeneracy $n$ of a level is an integer number that can be either  even or odd. By writing $n$ as $(2j+1)$, the $j$ value associated with odd  $n$  is an integer while for even $n$, it is a half-integer. The simplest way to label the corresponding $(2j+1)$ states then is through $m$ with $m=(j,j-1, \cdots,-j)$. So, for $n=2$, the states are labeled as $(\pm1/2)$; for $n=3$, the states are labeled as $(\pm1,0)$; for $n=4$, the states are labeled as $(\pm3/2,\pm1/2)$, and so on... 
 
This understanding of $(j,m)$ indices actually is quite general. It is also valid for  the degenerate levels of atomic electrons, even if they are commonly understood differently: indeed, due to the spherical symmetry of the potential felt by atomic electrons, these electrons have an \textit{orbital} motion  associated with an \textit{orbital angular} momentum  operator $\textbf{L}$ with integer eigenvalues denoted as $l$,  the corresponding $L_z$ eigenstates having an odd degeneracy $(2l+1)$ and a parity $(-1)^l$. A smart angular momentum $\textbf{J}=\textbf{L}+\textbf{S}$ formalism, that mixes orbital and spin subspaces, has been constructed. The angular momentum $\textbf{J}$ has half-integer eigenvalues denoted as $j$,  the corresponding $J_z$ eigenstates  having an even degeneracy $(2j+1)$. 

Yet, this $\textbf{J}=\textbf{L}+\textbf{S}$ formalism used in atomic physics does not hold for electrons in a periodic semiconductor crystal because electrons do not have the orbital motion induced by the atomic central potential. Still, the atomic notations are mathematically convenient for a cubic crystal, provided that $(x,y,z)$ are taken along the lattice axes while for atoms, these axes can be chosen at will, due to the central symmetry of the problem.

\end{document}